\documentclass[twocolumn,showpacs,aps,prl]{revtex4}% Physical Review Letter

\usepackage{graphicx}  % needed for figures
\usepackage{dcolumn}   % needed for some tables
\usepackage{bm}        % for math
\usepackage{amssymb}   % for math

\usepackage{verbatim}

\begin{document}

\title{Non-Kondo zero-bias anomaly in quantum wires}

\author{T.-M. Chen}
\thanks{These authors contributed equally to this work}
\author{A. C. Graham}
\thanks{These authors contributed equally to this work}
\author{M. Pepper}
\author{I. Farrer}
\author{D. A. Ritchie}

\affiliation{
Cavendish Laboratory, J J Thomson Avenue, Cambridge CB3 0HE, United Kingdom
}

\date{\today}

\begin{abstract}

It has been suggested that a zero-bias conductance peak in quantum
wires signifies the presence of Kondo spin-correlations, which might
also relate to an intriguing one-dimensional (1D) spin effect known as the 0.7
structure.  These zero-bias anomalies (ZBA) are strongly
temperature dependent, and have been observed to split into two
peaks in magnetic field, both signatures of Kondo correlations in
quantum dots. We present data in which ZBAs in general do not split as magnetic field is increased up to 10~T. A few of our ZBAs split in magnetic field but by significantly \textit{less} than the Kondo splitting value, and evolve back to a single peak upon moving the 1D constriction laterally. The ZBA therefore does not appear to have a Kondo origin, and instead we propose a simple phenomenological model to reproduce the ZBA which is in agreement mostly with observed characteristics.

\end{abstract}

\pacs{73.21.Hb, 73.23.Ad, 72.10.Fk}
\maketitle

 The varied manifestations of the Kondo effect \cite{kondo}, in
 systems as diverse as carbon nanotubes \cite{cobden},
 semiconductor quantum dots \cite{goldhaberN}, two-dimensional (2D) molecular
 systems \cite{iancu} and low-density mesoscopic 2DEGs \cite{arindam},
 have continued to interest experimentalists and theoreticians
 alike. Hence, when the Kondo effect was suggested as a possible explanation \cite{marcus}
 of a many-body phenomenon in quantum
 wires known as the 0.7 structure \cite{thomas96}, this
 created great interest in the physics of quantum wires. A microscopic
 explanation of the 0.7 structure has proved to be a difficult
 theoretical challenge
 \cite{spivak,bruus,klironomos,meir,wang96,karl05,lassl}.
 The suggested link with Kondo physics answered some questions, such as
 why the 0.7 structure rises in conductance and disappears
 with decreasing temperature \cite{kristensenprb,thomas96}, but
 raised others: could Kondo physics possibly occur in an open system\cite{meir06,Ihnatsenka07}?
 Why does a dc-bias strengthen the 0.7 structure and increase its conductance \cite{kristensenprb,patel91b}, when Kondo correlations are quickly destroyed by finite bias \cite{meir93}? Very recently, a channel with a variable open dot geometry shows that the Kondo effect and the 0.7 structure are completely separate\cite{Francois08}. It was also found that the 0.7 structure is not associated with backscattering of electron waves and reduction in transmission probability\cite{Czapkiewicz}. As more and more studies imply the Kondo effect is not linked to the 0.7 structure, this topic requires further studies.

  The key evidence linking the 0.7 structure and the Kondo effect,
  is the presence of a `zero-bias anomaly' (ZBA), or conductance peak at
  zero dc-bias, in the region of the 0.7 structure \cite{marcus};
  the Kondo effect causes a similar feature in quantum dots
  \cite{goldhaberN}.  This ZBA was found to broaden
  and disappear with increasing temperature, and to split in two
  with increasing magnetic field, both of which can be interpreted as manifestations of the Kondo effect.

Here, in contrast, we present evidences that ZBA characteristics in
quantum wires are inconsistent with spin-one-half Kondo physics. We observe the single ZBA peak up to an in-plane magnetic field of $10~$T, and at conductances less than $e^2/h$ such that only a single spin-type can be present. Such a fully spin-polarised regime prevents the spin-flips occurring which are essential to the spin-one-half Kondo effect. Thus the measured ZBA does not relate to Kondo effect as it is blocked in this regime. We have also found a few ZBAs splitting in a magnetic field, but all of these evolve back to a single peak when the 1D channel is laterally shifted. It implies that the splitting of the ZBA is related to disorder rather than Kondo splitting. We propose instead a simple phenomenological model whereby small increases in the 1D subband energies as a function of dc bias can directly reproduce temperature-dependent ZBAs.

Our split-gate devices are fabricated on a GaAs/Al$_{0.33}$Ga$_{0.67}$As
heterostructure, whose two-dimensional electron gas formed at
96~nm below the surface, has a mobility of $8.04 \times
10^5$~cm$^2$/Vs and a carrier density of $1.44 \times
10^{11}$~cm$^{-2}$ in the dark. All the devices used in this work
have a width of 0.8~$\mu$m, and lengths from 0.3~$\mu$m to
1.5~$\mu$m and were measured in a dilution refrigerator at $60~$mK.

%insert Fig.1 %
\begin{figure}[t!]
\begin{center}
\includegraphics[width=0.95\columnwidth]{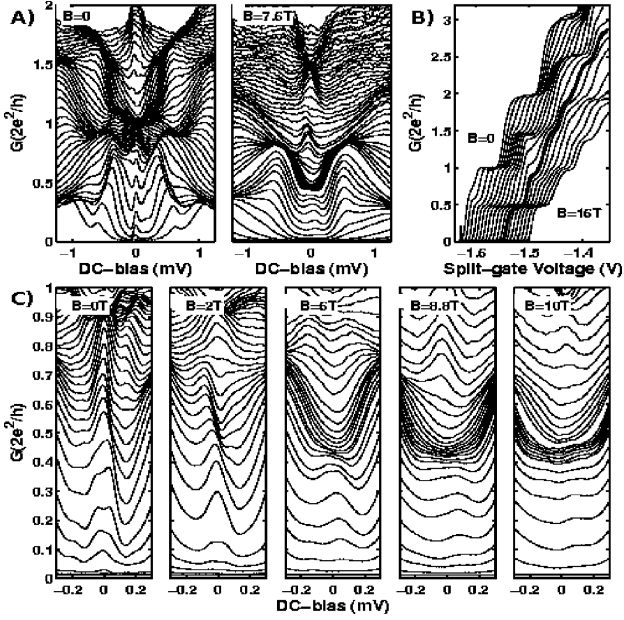}
\end{center}
\caption[Figure1]
{\small \textbf{\textit{A)}} Differential conductance vs dc-bias
for fixed split-gate voltages, at $B=0$ and $B=7.6~$T, both showing a
clear zero-bias anomaly (ZBA).
\textbf{\textit{B)}} Differential conductance vs split-gate
voltage for $B=0\rightarrow16~$T. \textbf{\textit{C)}} DC-bias
data showing the evolution of the ZBA with
increasing field.}
\label{fig1}
\end{figure}
%continued%

Typical differential conductance data exhibiting a strong
zero-bias anomaly and 0.7 structure are shown in
Fig.~\ref{fig1}(a).  With increasing magnetic field $B$, the ZBA weakens, but does not
split in two, and a single peak is still visible at $B=10$~T,
as shown by the more detailed evolution in Fig.~\ref{fig1}(c). A ZBA
of Kondo origin should have been split by $|2g^*\mu_{B}B|=1.2~$meV
at $B=10$~T \cite{meir93,Costi}, where the effective \textit{g}-factor $|g^*|=1.06$ is obtained from the spin gap measured by dc-bias spectroscopy as well as the technique reported in Ref.~\cite{patel91b}. Note that the \textit{g}-factor of $1.06$ measured in this work is consistent with the \textit{g} value reported in previous 1D studies\cite{thomas96,marcus}. Furthermore, Fig.~\ref{fig1}(c) ($B=10~$T) exhibits a strong bunching of traces at $e^2/h$ (i.e., a plateau) indicating complete spin-polarisation which would prevent Kondo spin-flip scattering from occurring. The 0.7 analog
and the weakening $1.5\times(2e^2/h)$ plateau at $B=9.6~$T (bold trace)
in Fig.~\ref{fig1}(b) indicates that a crossing of spin-split
subbands has already occurred by $B=10T$, further implying a
completely spin-polarised system\cite{abiprl}. Thus, the evidence that the ZBA still occurs in a completely spin-polarised regime, in which the Kondo effect is blocked, and does not split in magnetic field implies that Kondo is unlikely to be the cause of the ZBA in quantum wires.

%insert Fig.2 %
\begin{figure}[t!]
\begin{center}
\includegraphics[width=0.95\columnwidth]{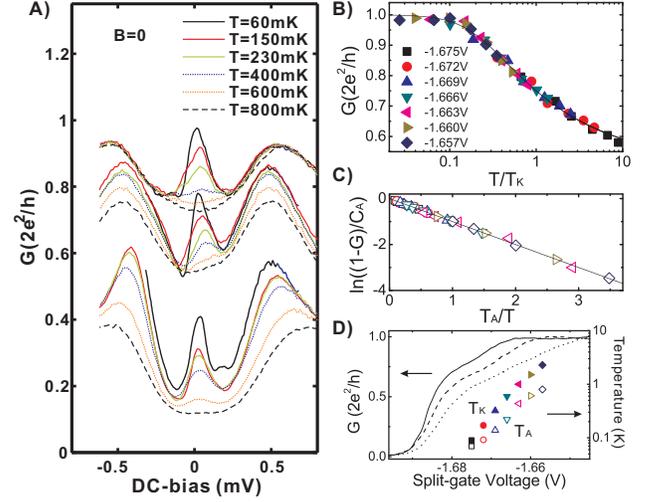}
\end{center}
\caption[Figure2] {\small (Color online) \textbf{\textit{A)}} Temperature dependence of differential conductance vs dc-bias at $B=0$. \textbf{\textit{B)}} Linear conductance as a function of scaled temperature $T/T_K$ at various $V_g$, where the Kondo temperature $T_K$ is the fit parameter in the modified Kondo function. \textbf{\textit{C)}} The same experimental data shown in an Arrhenius plot: $ln((1-G(T))/C_A)$ as a function of scaled temperature $T_A/T$, where the activation temperature $T_A$ and $C_A$ are the fit parameters in Arrhenius function. \textbf{\textit{D)}} $T_K$(solid symbols) and $T_A$(open symbols) extracted from the fits of both models, along with the conductance at $60$~mK (solid), $230$~mK (dashed), and $800$~mK (dotted).} \label{fig2}
\end{figure}
%continued%

Another argument to link the ZBA with the Kondo state is the Kondo-like temperature dependence of the equilibrium conductance\cite{marcus}. However, this equilibrium conductance in the steeply-rising region in 1D systems is also expected to have an activated temperature dependence\cite{kristensenprb}. The ZBAs weaken and disappear with increasing temperature in both zero (Fig.\ref{fig2}A) and finite magnetic fields (not shown). Figures~\ref{fig2}(b)\&(c) show that the temperature dependence of the ZBAs is well described by both the empirically modified Kondo function\cite{marcus}, $G_K=(2e^2/h)[0.5(1+(2^{1/s}-1)(T/T_K)^2)^{-s}+0.5]$ with $s=0.22$, and the activation model\cite{kristensenprb}, $G_A=(2e^2/h)[1-C_Ae^{-T_A/T}]$. Note that this modified Kondo function can only describe the ZBA between $0.5(2e^2/h)$ and $1\times(2e^2/h)$, since the lower limit of $G_K$ equals $0.5(2e^2/h)$; however, in experiments the ZBA is observed continuously from $0$ to $2e^2/h$. The Kondo temperature, $T_K$, and the activation temperature, $T_A$, extracted from the fits of these two functions are furthermore shown in Fig.~\ref{fig2}(d). The values of $T_K$ measured for different samples in this work and previous study\cite{marcus} are all close to each other, and also rule out a possible reason for the disappearance of the splitting of ZBAs, i.e., $k_B T_K > g^* \mu B$ \cite{Costi}. Note that the equivalence of Kondo-like and activated fitting of the same data was also found by Cronenwett \textit{et al.} (Footnote 25 in Ref.\cite{marcus}, and Ref.\cite{cronenwettthesis}). It appears that the temperature dependence of the ZBA does not favor the Kondo model over the activation model. It is worth emphasizing that the ZBAs reported here and also in previous studies\cite{marcus,cronenwettthesis} can not be fitted to the usual Kondo function \cite{gordon_PRL98}.

%insert Fig.3 %
\begin{figure}[t!]
\begin{center}
\includegraphics[width=0.9\columnwidth]{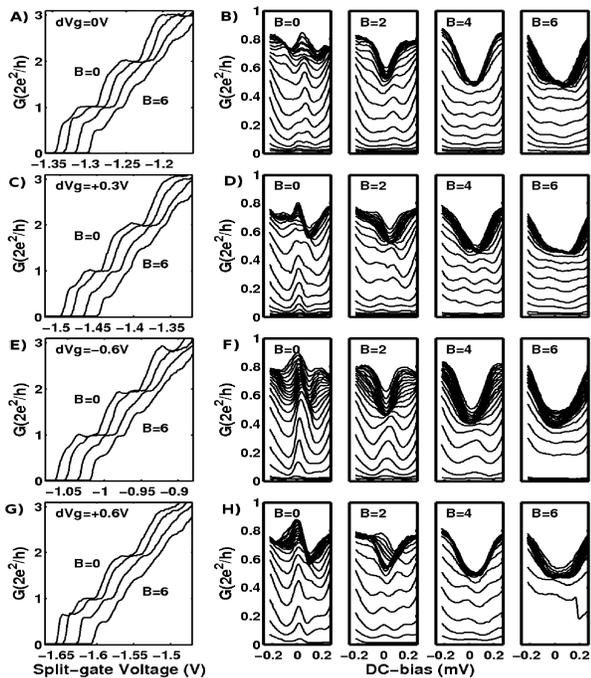}
\end{center}
\caption[Figure3] {\small \textbf{\textit{A)}} Differential
conductance vs split-gate voltage at $B=0,2,4$ and $6~$T.
\textbf{\textit{B)}} Differential conductance vs dc-bias for fixed
split-gate voltages, at $B=0,2,4$ and $6~$T, showing a clear
ZBA which splits into two with increasing $B$.
\textbf{\textit{C)}}, \textbf{\textit{D)}} Same as A) and B) but
with a difference of $dV_g=+0.3~$V between the voltages on the two
split-gates --- the splitting of the ZBA with
increasing $B$ is even clearer. \textbf{\textit{E)}},
\textbf{\textit{F)}} Same as A) and B) but for
$dV_g=-0.6~$V --- the ZBA is stronger for all $B$, but no longer exhibits any
splitting.\textbf{\textit{G)}},
\textbf{\textit{H)}} Same as A) and B) but for
$dV_g=+0.6~$V --- these data are very different to E) and F), indicating that disorder strongly affects ZBAs.} \label{fig3}
\end{figure}
%continued%

Although most of our samples show only a single
peak at all $B$, two samples do exhibit splitting of the
ZBA into two peaks with increasing $B$
[Fig.\ref{fig3}(b), with conductance vs gate voltage characteristics
in (a)]. However, the peaks are split by $0.17~$meV at $B=4~$T, which is only one-third of the value expected for Kondo splitting, $2g^*\mu_{B}B=0.48~$meV. It was suggested that the Kondo splitting could be less than $|2g^*\mu_{B}B|$ and have a minimum value of about $|\frac{4}{3}g^*\mu_{B}B|$ due to the interaction between the impurity spin and the lead electrons\cite{Moore}. However, the splitting of $0.17~$meV at $B=4$T is still less than the minimum value $\frac{4}{3}g^*\mu_{B}B=0.32~$meV for a Kondo state.

The strength of the split peaks can be enhanced by shifting the quantum wire laterally through the 2DEG
\cite{williamson}, by setting a difference between the voltages on each split-gate of $dV_g=+0.3~$V [Fig.~\ref{fig3}(d)]. However, if the wire is shifted in the other direction, by applying a difference
of $-0.3~$V, then the form of the ZBA changes to an asymmetric
single peak (not shown), which becomes almost symmetric and very
pronounced at $dV_g=-0.6~$V [Fig.~\ref{fig3}(f)]---in Fig.~\ref{fig3}(f) at $B=4~$T, the ZBA is a single peak, clearly not two broad peaks separated by $2g^*\mu_{B}B=0.48~$meV. In contrast, the observed ZBA which does not split with magnetic fields at $dV_g=0$ remains as a single peak when the wire is shifted laterally.

The difference in our data for $dV_g=+0.6~$V [Fig.~\ref{fig3}(f)],
compared to $dV_g=-0.6~$V [Fig.~\ref{fig3}(h)], is evidence that the ZBA
is strongly affected by a disordered confining potential, which
changes as the wire is moved laterally; in the absence of disorder,
the conductance characteristics should be the same for $dV_g=\pm
0.6~$V. The recurring of a symmetric single ZBA peak from a split ZBA by laterally moving the 1D channel clearly suggests that disorder is related to the splitting of the ZBA in a magnetic field, as opposed to Zeeman splitting of a Kondo resonance.

It has been shown that the differential conductance of an asymmetric electric constriction, caused by disorder in our case, is strongly affected by magnetic field and bias voltage \cite{LofgrenPRL04}. Moreover, impurity causes resonant interference of the electrons in the wire; electron interference is also expected to change with dc bias and magnetic field, as diamagnetic shift, Zeeman splitting and localisation can all alter the energy spectrum. All these provide a reason why zero-bias anomalies in a few devices exhibit splitting-like behaviour with magnetic field.

\begin{figure}[t!]
\begin{center}
\includegraphics[width=0.9\columnwidth]{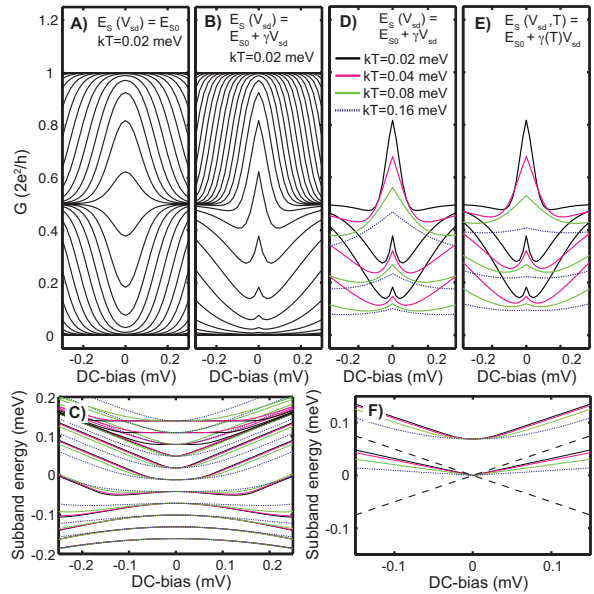}
\end{center}
\caption[Figure4]
{\small (Color online) \textbf{\textit{A)}} Calculated differential conductance vs dc-bias,
assuming subband energy, $\rm E_{s}=E_{s0}$, is fixed w.r.t. the
average of the source and drain chemical potentials, where $\rm
E_{s0}$ is the subband energy at zero dc-bias.
\textbf{\textit{B)}} Differential conductance vs dc-bias,
calculated in the same way, but including a linear increase,
$\rm E(V_{sd})$, in the subband energy $\rm E_{s}$, as a function
of dc-bias. A zero-bias anomaly is present. \textbf{\textit{C)}}Variation in subband energy required to keep density fixed as function of DC-bias, for $kT=0.02-0.16~$meV. Curves at $y>0$ $(y<0)$ are for the subband sitting above (below) $E_f$ at $V_{sd}=0$.
\textbf{\textit{D)}}The temperature variation of the
characteristics in B). \textbf{\textit{E)}} Same as D), but now
using a linear increase $\rm E(V_{sd},T)$ in subband energy, which
decreases with increasing temperature. \textbf{\textit{F)}} Comparison of subband energy used in E) to calculate $dI/dV_{sd}$ (bottom) to curves for fixed density (top) taken from C). }
\label{fig4}
\end{figure}

Some simple simulations were performed to demonstrate how a ZBA might occur using a very general phenomenology, which does not include spin. Figure~\ref{fig4}(a) shows differential conductance characteristics assuming subband energies ${\rm E_s}$ stay fixed as a function of dc-bias\cite{Moreno}, with respect to the average of the source and drain
chemical potentials $(\mu_s+\mu_d)/2$ --- as expected, there is no ZBA. In contrast, Fig.~\ref{fig4}(b) shows characteristics where the subband energy rises linearly at a small rate ($\gamma=0.33~$meV per mV) with
increasing dc-bias (${\rm E_s(V_{sd})=E_s(0)+\gamma
V_{sd}}$) --- a sharp ZBA is now present.  We
have used a linear increase in energy with $V_{sd}$ for
simplicity, but other increasing functions of $V_{sd}$ also give
ZBAs.  Thus,  a zero-bias anomaly occurs in 1D whenever the subband energy rises slightly with increasing dc-bias,
irrespective of the precise functional form of this rise in energy.

There is a fundamental reason for our phenomenological model wherein the subband energy rises upward with increasing dc-bias. In 1D systems, a fixed subband energy, with respect to $(\mu_s+\mu_d)/2$, ensures that 1D density must change as a function of the square root of dc bias, which is unlikely to be energetically favourable because of the cost in Coulomb energy. Therefore, simply minimizing the Coulomb energy alone gives a reason why subband energy should change with dc-bias. The curves in Fig.~\ref{fig4}(c) show how the 1D subband energy must vary in order to keep the 1D density constant with increasing dc bias. It further suggests an upward shift rate of $\gamma=0.33~$meV/mV when the subband is near the chemical potential to give rise ZBAs. Note that completely fixed density is not required to reproduce ZBAs; simply reducing changes in density also causes subbands to increase in energy with $V_{sd}$, giving a ZBA as well.

Although the zero-bias anomalies arising from a linear increase in subband energy with $V_{sd}$ broaden and weaken with increasing temperature [Fig.~\ref{fig4}(d)], for the ZBA to disappear completely at high temperatures, as in experiment, it is necessary that the linear increase in subband energy be temperature dependent, tending to zero at high
temperatures, as shown in Fig.~\ref{fig4}(e) --- $\gamma=0.33~$meV per mV at the lowest temperature ($k_{B}T=0.02~$meV) gives a strong narrow ZBA, which broadens and disappears at the highest temperature ($k_{B}T=0.16~$meV) for which $\gamma=0.1~$meV/mV.

There is a fundamental justification for introducing this temperature-dependent
correction to give the correct phenomenology. Figure~\ref{fig4}(c) clearly demonstrates that the variations in subband energy with $V_{sd}$ due to conservation of density are temperature-dependent and disappear with increasing temperature, just as is required in our phenomenological model [Fig.~\ref{fig4}(e)] in order to reproduce a ZBA that disappears at high $T$. Fig.~\ref{fig4}(f) shows that the energy shifts used to give a ZBA which disappears at high $T$ in our phenomenological model (lower curves), are similar in magnitude to those required to keep 1D density fixed (upper curves).

Here we do not claim that complete or partial conservation of density is necessarily the cause of the experimentally observed ZBAs, as our simple model is only of qualitative significance. Since Coulomb and exchange interactions vary with density, they are also expected to have an impact on the shift of subband energy with $V_{sd}$ and require further investigations. It is important to stress that this simple model demonstrates that the ZBAs in quantum wires, unlike those in quantum dots, do not need to have the Kondo mechanism as the nonlinear conductance in 1D changes easily when subband energy varies with source-drain bias. This could provide a basis for future theoretical study in anomalous nonlinear conductance features, such as the ZBA and its two significant side peaks which cannot be explained by the Kondo model.

The model presented here, though simple, is consistent with the temperature characteristics of the ZBAs which appears to have an activated behaviour. In addition, it explains why ZBAs do not split with increasing magnetic fields and why ZBAs still occur in a completely spin-polarised phase in which the Kondo spin-flip is prohibited. However, the model cannot interpret the suppression of the ZBAs with increasing magnetic field and will need a field-dependent correction. A more sophisticated theoretical study with Coulomb and exchange-correlation interactions taken into account is thus highly needed.

To conclude, we have found zero-bias anomalies in general suppress with increasing magnetic fields without being accompanied with splitting. Some of ZBAs persist up to a very large field when only one spin species occupies the quantum wire, wherein the Kondo effect is not expected to occur. While two samples exhibit split ZBAs in magnetic fields, the splitting value is not consistent with the the Kondo model and, more importantly, the split ZBA evolves back into a single ZBA peak when the 1D constriction is moved laterally. The measurement of ZBAs when quantum wires are laterally shifted strongly suggests that the splitting of ZBAs is related to disorder, as opposed to Kondo splitting. A simple phenomenological model is suggested to explain how a ZBA can occur when the 1D subband energy rises with increasing source-drain bias, clearly showing that the Kondo mechanism is not required for zero-bias anomalies to occur in quantum wires.

We acknowledge useful discussions with F. Sfigakis, K. Das Gupta and K.-F. Berggren. This work was supported by EPSRC (UK). T.M.C. acknowledges an ORS award and financial support from the Cambridge Overseas Trust. A.C.G. acknowledges support from Emmanuel College, Cambridge.

%\bibliography{refs2}

\end{document}